\documentclass[10pt, hidelinks, onecolumn]{article}

\usepackage{color}
\usepackage[utf8]{inputenc}
\usepackage[T1]{fontenc}
\usepackage{textcomp}
\usepackage{graphicx}

\usepackage{amsmath}
\usepackage{amsfonts}
\usepackage{amssymb}
\usepackage{bbm}
\usepackage{nicefrac}
\usepackage{booktabs}
\usepackage{multirow}
\usepackage{hyperref}

\usepackage{algorithm}
\usepackage{algorithmic}
  \usepackage[caption=false,font=footnotesize]{subfig}
\usepackage{fixltx2e}
\usepackage{url}

\newtheorem{theorem}{Theorem}
\newtheorem{observation}[theorem]{Observation}%

\newcommand{\R}{\mathbb{R}}

\newcommand{\brac}[1]{\left(#1\right)}

\newcommand{\norm}[1]{\left\| #1 \right\|}

\newcommand{\alref}[1]{{Algorithm~\ref{#1}}}
\newcommand{\figref}[1]{{Figure~\ref{#1}}}
\newcommand{\tabref}[1]{{Table~\ref{#1}}}

\newcommand{\appref}[1]{{Appendix~\ref{#1}}}
\newcommand{\secref}[1]{{Section~\ref{#1}}}

\newcommand{\fbp}{\text{FBP}}
\newcommand{\nnfbp}{\text{NN-FBP}}
\newcommand{\mlp}{\text{MLP}}

\newcommand{\fp}{W}
\newcommand{\bp}{W^T}

\newcommand{\numpixels}{\ensuremath{N}}
\newcommand{\numangles}{\ensuremath{N_{\text{a}}}}
\newcommand{\numfilters}{\ensuremath{N_{\text{h}}}}
\newcommand{\filterwidth}{\ensuremath{N_{\text{f}}}}
\newcommand{\numbases}{\ensuremath{N_{e}}}
\newcommand{\numsplits}{\ensuremath{N_{\text{s}}}}
\newcommand{\numtrain}{\ensuremath{N_{\text{T}}}}

\newcommand{\projspace}{\ensuremath{\R^{m}}}
\newcommand{\recspace}{\ensuremath{\R^{n}}}
\newcommand{\filterspace}{\ensuremath{\R^{\filterwidth}}}

\newcommand{\h}{\mathbf{h}}

\renewcommand{\vec}[1]{\mathbf{#1}}
\newcommand{\x}{\vec{x}}
\newcommand{\y}{\vec{y}}

\newcommand{\xfbp}{\vec{x}_{\text{FBP}}}
\newcommand{\xfbpother}{\vec{x}_{\text{FBP}, l\neq{} j}}
\newcommand{\xfbpl}{\vec{x}_{\text{FBP}, l}}
\newcommand{\xfbpj}{\vec{x}_{\text{FBP}, j}}
\newcommand{\xtgt}{\vec{x}_{\text{Target}}}

\renewcommand{\a}{\vec{a}}
\renewcommand{\b}{\vec{b}}

\newcommand{\ee}{\vec{e}}

\newcommand{\net}{\ensuremath{}\mathrm{CNN}_{\param}}
\newcommand{\trainednet}{\ensuremath{}\mathrm{CNN}_{\paramhat}}
\newcommand{\param}{\ensuremath{\theta}}
\newcommand{\paramhat}{\ensuremath{\theta^{\star}}}

\newcommand{\address}[1]{#1}

\def\ead#1{\vspace*{5pt}\address{E-mail: \mailto{#1}}}
\def\mailto#1{{\tt #1}}

\def\submitto#1{\vspace{28pt plus 10pt minus 18pt}
     \noindent{\small\rm Submitted to: {\it #1}\par}}

\newcounter{questioncounter}  

\begin{document}
\title{Noise2Filter: fast, self-supervised learning and real-time reconstruction for 3D Computed Tomography}

\author{Marinus~J.~Lagerwerf$^{1, \dagger}$,
  Allard~A.~Hendriksen$^{1, \dagger}$\\
    Jan-Willem Buurlage$^{1}$,, and
  K.\ Joost Batenburg$^{1,2}$
}
\maketitle
\address{$^1$ Computational imaging group, Centrum Wiskunde \& Informatica, Amsterdam 1098 XG, The Netherlands}

\address{$^2$ Leiden Institute of Advanced Computer Science, Leiden Universiteit, 2333 CA Leiden, The Netherlands}

\address{$^\dagger$ Equal contribution.}

\ead{m.j.lagerwerf@cwi.nl}

\begin{abstract}

  At X-ray beamlines of synchrotron light sources, the achievable
  time-resolution for 3D tomographic imaging of the interior of an object has
  been reduced to a fraction of a second, enabling rapidly changing structures
  to be examined.
  The associated data acquisition rates require sizable computational resources
  for reconstruction.
  Therefore, full 3D reconstruction of the object is usually performed after the
  scan has completed.
  Quasi-3D reconstruction --- where several interactive 2D slices are computed
  instead of a 3D volume --- has been shown to be significantly more efficient,
  and can enable the real-time reconstruction and visualization of the interior.
  However, quasi-3D reconstruction relies on filtered backprojection type
  algorithms, which are typically sensitive to measurement noise.
  To overcome this issue, we propose Noise2Filter, a learned filter method that
  can be trained using only the measured data, and does not require any
  additional training data.
  This method combines quasi-3D reconstruction, learned filters, and
  self-supervised learning to derive a tomographic reconstruction method that
  can be trained in under a minute and evaluated in real-time.
  We show limited loss of accuracy compared to training with additional training
  data, and improved accuracy compared to standard filter-based methods.

\end{abstract}
\vspace{2pc}

\submitto{Machine Learning: Science and Technology}

\section{Introduction}\label{sec:introduction} Computed tomography is a
non-destructive imaging technique with applications in
biology~\cite{santos-2014-in-vivo}, energy research~\cite{xu-2020-optim-image},
materials science~\cite{garcia-moreno-2018-time}, and many other
fields~\cite{de-2018-tomob}.
In a tomographic scan, a rotating object is positioned between a source emitting
penetrating radiation and a detector that captures the projections of the
object.
Tomographic reconstruction algorithms compute a 3D image of the interior of the
object from its projections.
Besides extensive use in medical and laboratory settings, tomography is
routinely used at synchrotron facilities, where advances in the last decade have
enabled time-resolved imaging of the interior structure of a rapidly changing
object~\cite{santos-2014-in-vivo,xu-2020-optim-image,garcia-moreno-2018-time}.
So far, reconstruction algorithms are typically operated offline, enabling
visualization of the object only after a scan has completed.

Recent advances in tomographic reconstruction enable real-time interrogation of
the reconstructed volume during the scanning process using a quasi-3D reconstruction protocol~\cite{buurlage2018real,buurlage2019real}.
In this framework, arbitrarily oriented slices are selected for reconstruction
and can be interactively rotated and translated, after which they are
reconstructed and visualized virtually instantaneously.
This creates the illusion of having access to the full reconstructed 3D volume,
but at a fraction of the computational cost. 
The quasi-3D reconstruction protocol has been implemented in the RECAST3D software package.
The information gained from this quasi-3D visualization can be used to directly
steer the tomographic experiment, for instance, by adjusting an external
parameter --- such as temperature --- in response to changes in the interior of
the object.
In addition, the object can be re-positioned, or other acquisition parameters
can be adjusted to facilitate the best possible
reconstruction~\cite{vanrompay2020real}.

\begin{figure}
  \centering
  \subfloat[FBP reconstruction]{\includegraphics[width=0.45\textwidth]{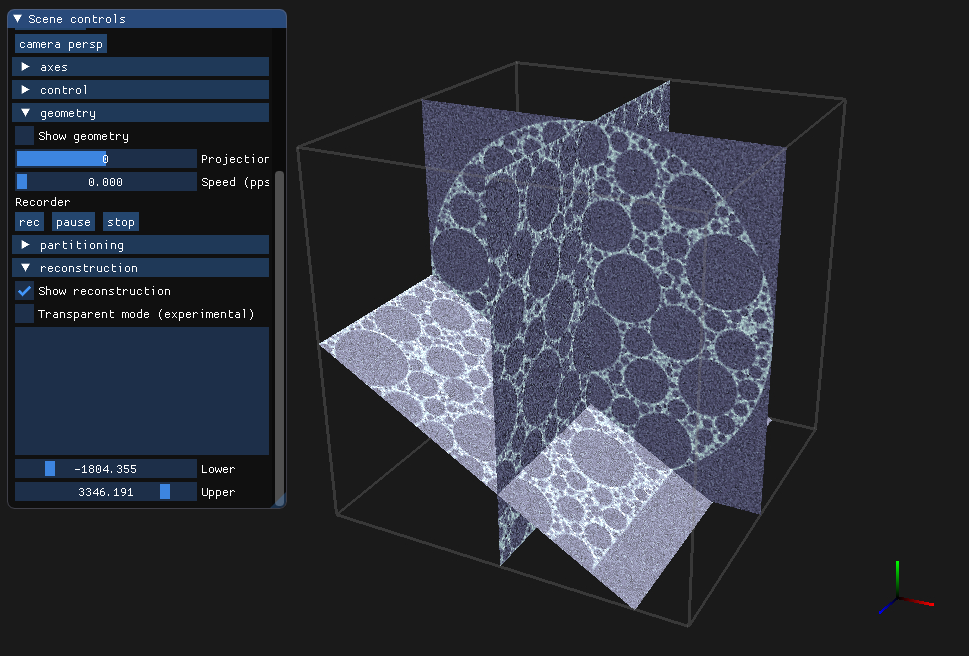}}
  \quad
  \subfloat[Noise2Filter reconstruction]{\includegraphics[width=0.45\textwidth]{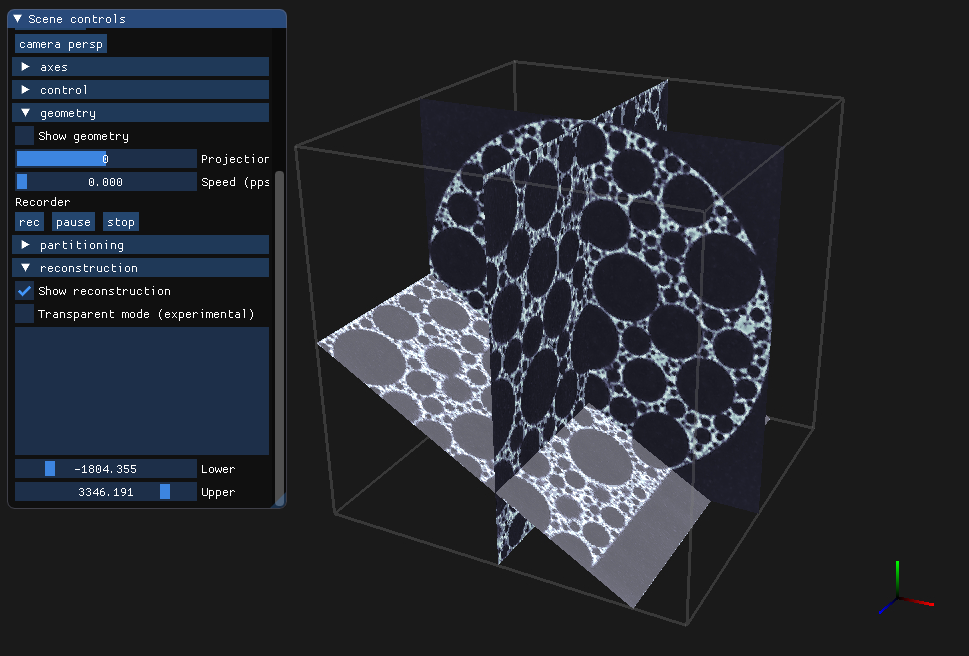}}
  \caption{\label{fig:recast3d-example} Real-time reconstructions using FBP and
    Noise2Filter of a high-noise acquisition using the RECAST3D software
    package.
    The highlighted slice is currently being moved.
  }
\end{figure}

Real-time 3D reconstruction is computationally demanding and data sizes are
substantial --- data acquisition rates of 7.7GB per second are not
uncommon~\cite{buurlage2019real}.
To attain real-time visualization, the quasi-3D reconstruction protocol is essentially limited to filtered
backprojection type methods, since it exploits the locality of backprojection to
obtain fast reconstructions.
Filtered backprojection (FBP) methods are sensitive to measurement noise,
leading to errors in the reconstructed slices~\cite{buzug-2008-comput-tomog}.
Therefore, application of these methods in the quasi-3D reconstruction protocol is not
well-suited to high-noise
acquisitions~\cite{pelt-2018-improv-tomog,xu-2020-optim-image}, as illustrated in
\figref{fig:recast3d-example}\textbf{.a}. 

In this paper, we combine a learning-based filtered reconstruction method with a
self-supervised training strategy to obtain Noise2Filter, a denoising FBP-type
reconstruction algorithm that can be applied in a quasi-3D reconstruction protocol.
This algorithm is designed to be both fast to train and
fast to evaluate.
Moreover, no additional training data is required other than the measured
projection data.

For dynamic scans, our method enables a possible use case where a static scan is
performed --- with the exact same acquisition rates as the dynamic scan ---
permitting the Noise2Filter method to be trained immediately.
After training for tens of seconds, real-time visualization of the dynamic
experiment can ensue, as illustrated in
\figref{fig:recast3d-example}\textbf{.b}.
In addition, we note that Noise2Filter can be used as a stand-alone reconstruction
method.

The first component of our method is the Neural Network filtered backprojection (NN-FBP) method~\cite{pelt-2013-fast-tomog}.
This method learns a set of filters, along with additional weights, and then forms the reconstructed image as a non-linear function of the individual FBP reconstructions, resulting in higher image quality than standard FBP.
However, its application requires the availability of ground truth or
high-quality reconstructed images.

This limitation can be overcome using the second component of our method,
Noise2Inverse~\cite{hendriksen-2020-noise2}, which is a recent machine learning
method designed to train denoising convolutional neural networks (CNNs) in
inverse problems in imaging.
To train a denoising CNN, the method splits the measured projection data to
obtain multiple statistically independent reconstructed slices, which are
presented to the network during training, without requiring additional high-quality data.

Our main contribution is that we show how to combine the NN-FBP method with the Noise2Inverse training strategy.
In addition, we demonstrate that NN-FBP training can be substantially
accelerated as compared to previous
methods~\cite{pelt-2013-fast-tomog}.
We evaluate our method on both simulated and experimental datasets, comparing to
both conventional filter-based methods and supervised NN-FBP.
Finally, we demonstrate that the method can be used in a quasi-3D reconstruction
protocol, and exhibit its potential use for dynamic control of tomographic
experiments.

The paper is structured as follows.
In the next section, we introduce the tomographic reconstruction problem and
the filtered backprojection algorithm.
In addition, we introduce quasi-3D reconstruction, NN-FBP, and Noise2Inverse.
These methods are combined in Section~\ref{sec:noise2filter-method}, where we
describe the Noise2Filter method.
In Sections~\ref{sec:experimental-setup} and~\ref{sec:experiments-results}, we
describe experiments to analyze the reconstruction accuracy of
Noise2Filter on real and simulated CT datasets.
Moreover, we study the hyper-parameters of the proposed method.
We discuss these results in Section~\ref{sec:conclusion}.

\section{Preliminaries}
\subsection{Reconstruction problem}
In parallel-beam tomography, an unknown object rotates with respect to a
planar detector and a parallel source beam.
Projections are acquired at a finite number $\numangles$ of rotation angles,
yielding 2D images defined on an $\numpixels \times \numpixels$ pixel grid.
The reconstruction problem can be modeled by a system of linear equations
\begin{align}
  \label{eq:IP}
  \fp \x =\y,
\end{align}
where the vector $\x\in \recspace$ denotes the unknown object,
$\y\in \projspace$ describes the measured projection data, and \(\fp =(w_{ij})\)
is an \(m \times n\) matrix where \(w_{ij}\) denotes the contribution of object
voxel \(j\) to detector pixel \(i\).
For the sake of simplicity we assume that the volume consists of
$n = \numpixels \times \numpixels \times \numpixels$ voxels, and the projection
dataset contains $m=\numangles \times \numpixels \times \numpixels$ pixels.

\subsection{Filtered backprojection methods}

We consider the filtered backprojection (FBP) method for parallel beam
tomography~\cite{natterer-2001-mathem-method}.
The FBP algorithm is a two step algorithm.
First, the data $\y\in \projspace$ is convolved over the width of the detector
with a one-dimensional filter $\h\in \filterspace$.
Next, the \emph{backprojection} $\bp : \projspace \rightarrow \recspace$ is
applied to compute a reconstruction $\xfbp \in \recspace$.
Expressing the FBP algorithm in terms of $\h$, $\y$ and $W$ yields
\begin{align}
  \label{eq:FBP}
  \fbp(\y, \h) = \bp (\y \ast \h) = \xfbp.
\end{align}

\begin{observation}[\bf FBP is two-step]\label{obs:two-step}
  The FBP algorithm consists of a \emph{filtering step} and a
  \emph{backprojection step}, and both can be computed separately.
  That is, the filtering can be performed in advance, and the backprojection can
  occur on demand.
  This technique will be used throughout the paper.
\end{observation}

We observe that the FBP algorithm can be described by a linear operator
when fixing either $\y$ or $\h$.
This will be exploited in the discussion of learned filter methods in
Section~\ref{sec:nn-fbp}.

\subsection{Quasi-3D reconstruction}\label{sec:recast}
A property shared by filtered-backprojection type algorithms is that they are
\emph{local}, in the sense that each voxel of the reconstructed volume
can be computed directly from the filtered data by
backprojecting onto only that voxel~\cite{buurlage2018real}.
Therefore, if one is interested in a subset of the reconstructed
volume, much of the computational cost of a full 3D reconstruction can be
avoided.
Specifically, if the reconstructed subset is a rectangular box or a slice,
efficient backprojection algorithms such as those implemented in the ASTRA
toolbox~\cite{van2016fast} can be used.
This reduces the computational cost of the backprojection step by an order~of~$N$.

\begin{observation}[\bf Locality]\label{obs:locality}
  The backprojection operator is \emph{local}.
  Computing the backprojection for a
  single voxel or a subset of voxels is therefore substantially faster than computing the backprojection for all voxels.
\end{observation}

This methodology has been implemented in the RECAST3D software
package~\cite{buurlage2018real}, which exposes a limited number of arbitrarily
oriented 2D slices.
These slices are interactive and can be manipulated by the technician of the
tomographic experiment.
This technique for real-time visualization has been successfully applied in
practice to acquisitions in micro-CT systems~\cite{coban2020explorative},
synchrotron tomography~\cite{buurlage2019real}, and electron
tomography~\cite{vanrompay2020real}.

\subsection{NN-FBP reconstruction algorithm}\label{sec:nn-fbp}

The NN-FBP algorithm learns a set of suitable filters and a set of weights, and
then forms a non-linear model that combines the individual FBP reconstructions.
The algorithm may be considered as a multi-layer
perceptron~\cite{hastie-2009-elemen-statis-learn} that operates pointwise on a
collection of suitable reconstructions. A schematic representation of the NN-FBP algorithm is given \figref{fig:nn-fbp}, a mathematical description is given below.

\begin{figure}[t!]
  \centering
  \includegraphics[width=\textwidth]{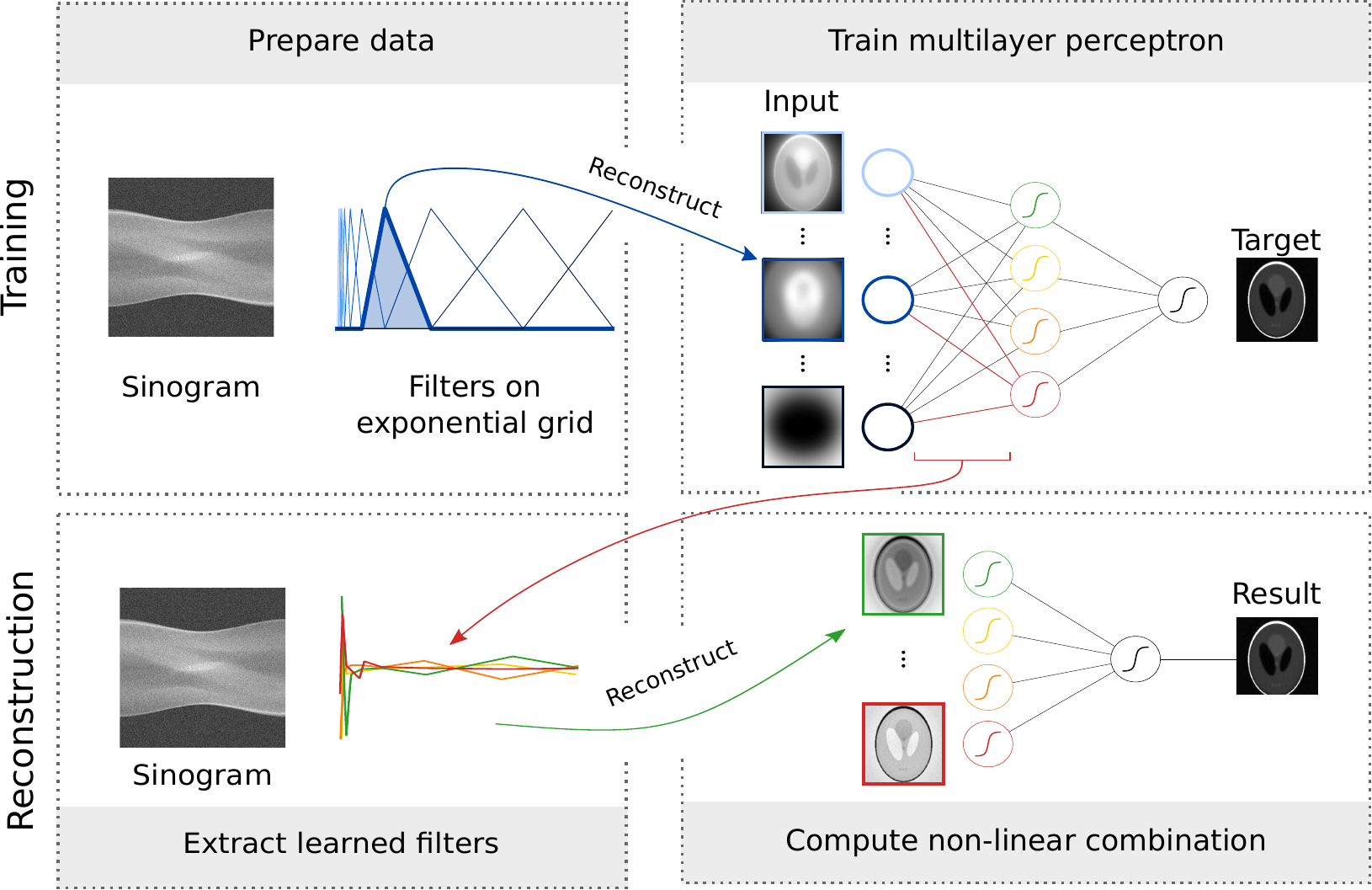}
  \caption{\label{fig:nn-fbp} An illustration of the NN-FBP method as applied to
    a noisy 2D Shepp-Logan sinogram.
    Before training, the data is reconstructed with filters
    $\ee_{1}, \ldots, \ee_{\numbases}$, defined on an exponential grid.
    These reconstructions $\x_{\ee_{1}}, \ldots, \x_{\ee_{\numbases}}$ are used
    as input for training a multilayer perceptron, as described in
    Equation~\eqref{eq:mlp}.
    The training target is a high-quality reconstruction.
    For reconstruction, learned filters $\h^{1}, \ldots, \h^{\numfilters}$ are
    extracted from the network (as indicated by the red arrow).
    Reconstructions are computed using the learned filters, and a non-linear
    combination is computed, as described in Equation~\eqref{eq:mlp-is-nnfbp}.
  }
\end{figure}

To obtain these reconstructions, we first make some general observations: a
filter $\h$ can be seen as a vector in $\filterspace$, and the FBP method is
linear in the filter when fixing the measured projection data $\y$.
Therefore, an FBP reconstruction can be expressed as a linear combination in the
basis of the filter.
Let $\ee_{1}, \ldots, \ee_{N_\text{f}}$ be any basis for the space of filters
$\filterspace$, such as the standard basis.
Define the reconstruction of $\y$ filtered by a basis element $\ee_{i}$ as
\begin{align}
  \label{eq:linear-decomp-filter-basis}
  \x_{\ee_{i}} := \bp \left(\y \ast \ee_{i} \right).
\end{align}
Then we can write the FBP reconstruction as a linear combination of these
reconstructions
\begin{align}
  \label{eq:fbp-decomposed}
  \xfbp(\y, \h)
  &= \sum_{i=1}^{N_\text{f}} \h_{i} \x_{\ee_{i}}
    = \sum_{i=1}^{N_\text{f}}   \bp \left(\y \ast \h_{i} \ee_{i} \right)
    = \bp \left(\y \ast \h \right),
\end{align}
where $\h_{i}$ denotes the coordinate of the $i$th basis element $\ee_i$.

Given a set of $\numfilters$ filters $\h^{1}, \ldots, \h^{\numfilters}$, we can define a
multi-layer perceptron (MLP) with one hidden layer as a function of the
reconstructions $\x_{\ee_{1}}, \ldots, \x_{\ee_{N_\text{f}}}$
\begin{align}
  \label{eq:mlp}
  \mlp_{\param}(\x_{\ee_{1}}, \ldots, \x_{\ee_{N_\text{f}}})
  &= \sigma \left(
    \sum_{k=1}^{\numfilters} \a_{k} \sigma\left(
        \sum_{i=1}^{N_\text{f}} \h_{i}^{k} \x_{\ee_{i}} - \b_{k}
    \right)
    - \b_{0}
    \right),
\end{align}
where $\sigma$ is a non-linear activation function, such as the sigmoid.
The multi-layer perceptron has free parameters
$\param = (\a, \b, \h^{1}, \ldots, \h^{\numfilters})$.
Plugging Equation~\eqref{eq:fbp-decomposed} into Equation~\eqref{eq:mlp},
we obtain the NN-FBP reconstruction algorithm
\begin{align}
  \label{eq:mlp-is-nnfbp}
  \nnfbp_{\param}(\y)
  &= \sigma \left(
    \sum_{k=1}^{\numfilters} \a_{k} \sigma\left(
        \fbp(\y, \h^{k}) - \b_{k}
    \right)
    - \b_{0}
    \right),
\end{align}
which is amenable to fast, parallel computation because it is a non-linear
combination of FBP reconstructions.

\begin{observation}[\bf pointwise]\label{obs:pointwise}
  Note that the multi-layer perceptron operates \emph{point-wise} on the voxels
  of the reconstructed volumes.
  Therefore, a \emph{single} voxel can be computed without having to reconstruct
  other voxels.
  This observation connects to the observation of locality on
  Page~\pageref{obs:locality}, and will return several times in this paper.
\end{observation}

\emph{Supervised training}~\cite{hastie-2009-elemen-statis-learn} is used to
determine the free parameters of the MLP defined in Equation~\eqref{eq:mlp}.
The goal is to approximate a suitable \emph{target} reconstruction $\xtgt$ by
minimizing
\begin{align}
  \label{eq:nnfbp-training}
  \norm{\mlp_{\param}(\x_{\ee_{1}}, \ldots, \x_{\ee_{N_\text{f}}}) - \xtgt}_{2}^{2},
\end{align}
i.e., the mean square error with respect to the target reconstruction.

The size of the training problem in Equation~\eqref{eq:nnfbp-training} is
related to the number of reconstructed volumes
$\x_{\ee_{1}}, \ldots, \x_{\ee_{N_\text{f}}}$ and the size of these
reconstructions, which suggests two techniques that may be used to accelerate
training.
First, to reduce the number of reconstructions, the filter is expressed on an exponentially
binned grid, which grows logarithmically in the width of the filter.
Since the filter width is proportional to the number of pixels in each detector
row, we have $\numbases = O(\log \filterwidth) = O(\log \numpixels)$.
This technique yields suitable filter approximations, as observed
in~\cite{pelt-2013-fast-tomog,lagerwerf2020automated}.
Second, training may be accelerated by sampling \emph{a subset of voxels} on
which to minimize Equation~\eqref{eq:nnfbp-training}, rather than the full
volume.
Subsampling is possible because NN-FBP operates pointwise, as noted in
Observation~\ref{obs:pointwise}.

To summarize, we can split the NN-FBP algorithm into three parts,
namely: (1) \emph{data preparation}, where the input training data
$\x_{\ee_{1}},\ldots, \x_{\ee_{\numbases}}$ is
computed, (2) \emph{network training}, where the weights $\paramhat$ for the
network are determined using a supervised learning approach, and (3) the
\emph{reconstruction algorithm}, which is summarized in \alref{alg:NN-FBP}.

\begin{algorithm}[!t]
  \caption{NN-FBP reconstruction algorithm}
  \begin{algorithmic}[1]\label{alg:NN-FBP}
    \STATE{Given projections $\y$ and a set of parameters
      $\paramhat~:=~\brac{\a, \b, \h^{1}, \ldots, \h^{\numfilters}}$.}
    \STATE{Compute the FBP reconstruction using the learned filters:}\\
    \FOR{$k = \{1,2,..,N_h\}$}
    \STATE{$\x_{\h^{k}} = \fbp(\y, \h^{k})$}
    \ENDFOR
    \STATE{Compute a non-linear combination of these reconstructions:\\
      $\nnfbp_{\paramhat}(\y)=\sigma\brac{
        \sum_{k=1}^{\numfilters}\a_k  \sigma\brac{\x_{\h^k} - \b_k}- \b_0
      }$
    }
  \end{algorithmic}
\end{algorithm}

\subsection{Noise2Inverse training}\label{sec:nois-train}

Noise2inverse is a technique to train a convolutional neural network (CNN) to
denoise reconstructed images in a self-supervised
manner~\cite{hendriksen-2020-noise2}.
This means that no additional training data is required beyond the acquired
noisy measurements.
The key idea is change the training strategy by splitting the projection dataset into subsets, computing
sub-reconstructions with these subsets and train a neural network mapping one
sub-reconstruction to another.

First, the projection data is split into $\numsplits$ sub-datasets such that
projection images from successive angles are placed in different sub-datasets
$\y_{1}, \y_{2}, \ldots, \y_{\numsplits}$.
The network is trained to predict the reconstruction from one subset using the
reconstruction of the other subsets.
Training therefore aims to find the parameter $\paramhat$ that minimizes
\begin{align}
  \label{eq:1}
  \sum_{j=1}^{\numsplits} \norm{\net(\fbp(\y_{j})) - \fbp(\y_{l\neq j})}_{2}^{2},
\end{align}
where $\fbp(\y_{j})$ denotes the reconstruction from one subset of the data,
and $\fbp(\y_{l\neq j})$ denotes the FBP reconstruction of the remaining
subsets.
We observe that the FBP reconstruction of a projection dataset is the mean of
the FBP reconstruction of each projection image individually, which enables us
to obtain
\begin{align}
  \label{eq:fbp-superlinear}
  \fbp(\y_{l\neq j}) = \frac{1}{\numsplits - 1} \sum_{l\neq j} \fbp(\y_{l}).
\end{align}

Now the \emph{original training data} can be denoised by applying the trained
network to each subreconstruction individually and averaging to obtain
\begin{align}
  \label{eq:n2i-reconstruction}
  \x_{\text{N2I}} = \frac{1}{\numsplits} \sum_{i=1}^{\numsplits}\trainednet(\fbp(\y_{i})).
\end{align}

In the previous discussion, we have assumed that the target images are
reconstructed from more subsets than the input images.
As in~\cite{hendriksen-2020-noise2}, we call this the \emph{1:X} strategy.
A reverse \emph{X:1} training strategy is also possible.
Here, the target is a single subreconstruction and the input is reconstructed
from the remaining sub-datasets.

Note that convolutional neural networks take into account the surrounding
structure of a voxel, typically a 2D slice, and thus do not operate pointwise.
Therefore, these networks are are an example where
Observation~\ref{obs:pointwise} does not apply.

\section{Noise2Filter method}\label{sec:noise2filter-method}
Our proposed method combines the three ideas introduced in the previous section.
The NN-FBP method is trained on a single projection dataset using the
Noise2Inverse training strategy.
This enables fast reconstruction of arbitrarily oriented slices using the
NN-FBP reconstruction algorithm in a quasi-3D reconstruction protocol.
\bigskip

\begin{figure}
  \centering
  \includegraphics[width=\textwidth]{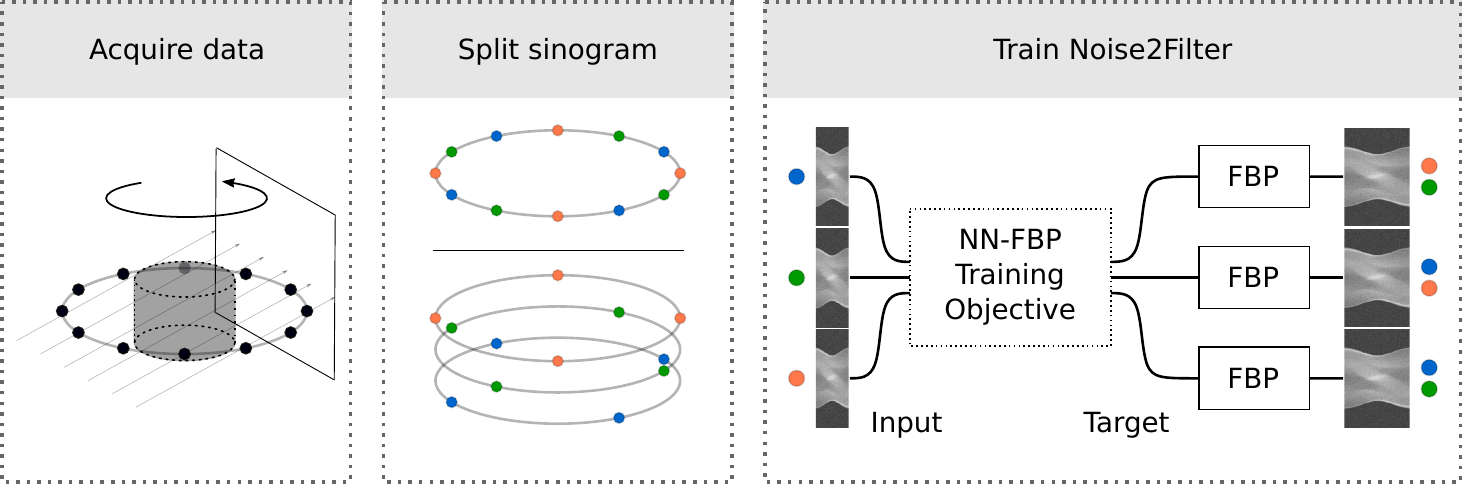}
  \caption{\label{fig:noise2filter-method} An illustration of the training of
    the Noise2Filter method.
    Data is acquired using a 3D parallel beam geometry.
    For each detector row, the sinogram is split in three sub-datasets such that
    acquisitions from successive projection angles belong to different
    sub-datasets.
    Each sub-dataset is used as input for NN-FBP training; the remaining
    sub-datasets are used in the target FBP reconstruction.
    This illustration depicts the \emph{1:X} strategy.
    In the \emph{X:1} strategy, the input is computed from the majority of the
    data, and the target from the minority, rather than vice versa.
  }
\end{figure}

\textbf{Training} The training procedure for the Noise2Filter method is similar
to the NN-FBP procedure described in~\cite{pelt-2013-fast-tomog}, with two
notable exceptions.
First,  instead of minimizing the supervised training objective in
Equation~\eqref{eq:nnfbp-training}, Noise2Filter minimizes a
self-supervised training objective similar to Equation~\eqref{eq:1}.
Second, training voxels are sampled from a subset of the reconstructed volume,
rather than the full volume.

As in Noise2Inverse, the projection data $\y$ is split into $\numsplits$ subdatasets with FBP reconstructions $\xfbpj, j=1, \ldots, \numsplits$.
For each subdataset $\y_{j}$, we denote with $\x_{j, \ee_{i}}$ a reconstruction
filtered with basis element $\ee_{i}$.

Training aims to minimize the difference between the MLP output of a subset of projection data and the FBP reconstruction of the remaining data.
For the \emph{1:X} training strategy, the MLP operates on a single subset of the
data and the target is reconstructed from the remaining subsets.
For the \emph{X:1} training strategy, on the other hand, the target is
reconstructed from a single subdataset, and the MLP operates on the remaining
subsets.
The self-supervised training objective thus becomes:
\begin{align}
  \label{eq:n2f-x1}
  \sum_{j=1}^{\numsplits} \norm{
  \mlp_{\param}(\x_{j, \ee_{1}}, \ldots, \x_{j, \ee_{\numbases}})
  - \xfbpother
  }_{2}^{2},
  &\quad(\text{\bf X:1 strategy})
  \\
  \label{eq:n2f-1x}
  \sum_{j=1}^{\numsplits} \norm{
  \mlp_{\param}(\x_{l\neq j, \ee_{1}}, \ldots, \x_{l \neq j, \ee_{\numbases}})
  - \xfbpj
  }^2_2,
  &\quad(\text{\bf 1:X strategy})
\end{align}
with
\begin{align}
  \xfbpother
  = \frac{1}{\numsplits - 1} \sum_{l \neq j} \xfbpl,
  \quad
  \x_{l\neq j, \ee_{i}}
  = \frac{1}{\numsplits - 1} \sum_{l \neq j} \x_{l, \ee_{i}} .
\end{align}
A schematic summary of the \emph{1:X} training strategy is given in \figref{fig:noise2filter-method}.

The second difference is related to the voxels that are considered for the
training.
Like NN-FBP, we minimize the training objective on a random sample of
$\numtrain$ voxels.
We have $\numtrain \ll \numpixels^{3}$, and increasing the sample size in
response to increasing object size has been observed to yield diminishing
returns.
Unlike NN-FBP, training voxels are sampled only from the reconstructions of the
axial, frontal, and longitudinal \emph{ortho-slices}, rather than the full
volume.
This choice substantially reduces the computational effort of the data
preparation step, as shown below.

\bigskip
\textbf{Data preparation} We discuss the \emph{1:X} strategy; similar statements
hold true for the \emph{X:1} strategy.

The data preparation step is the most computationally expensive part of the
method.
In this step, an input reconstruction $\x_{l\neq j, \ee_i}$ is computed for each
subdataset $\y_{j}$ and each basis element $\ee_{i}$.
In addition, a target reconstruction $\xfbpj$ is computed for each subdataset,
resulting in a total of $\numsplits (\numbases + 1)$ reconstructions.
These reconstructions are computed on the ortho-slices instead of the full
volume.
Due to locality --- see Observation~\ref{obs:locality} --- the computational
cost of the data preparation is therefore reduced by an order of $\numpixels$.

Note that the computational cost of the FBP algorithm scales linearly in the
number of projection angles, therefore the computational cost of this step is
equal to $3(\numbases + 1)$ FBP reconstructions of a 2D slice.
Splitting the projection data thus has no adverse effect on the performance.

\bigskip
\textbf{Reconstruction} The reconstruction algorithm is almost identical to the NN-FBP reconstruction algorithm
described in \alref{alg:NN-FBP}.
Whereas the aim of NN-FBP is to reconstruct the full volume, we aim only to
reconstruct slices on demand.
Therefore, reconstruction can be substantially accelerated.

We make use of Observation~\ref{obs:two-step} that the FBP algorithm can be
split in a filtering and backprojection step.
First, the acquired projection data is filtered with the learned filters and
cached.
Then, a single slice can be reconstructed using \alref{alg:NN-FBP}, which can
occur in real-time due to the locality of the backprojection
(Observation~\ref{obs:locality}) and the pointwise nature of the multi-layer
perceptron (Observation~\ref{obs:pointwise}).
Therefore, the reconstruction can be integrated in the quasi-3D reconstruction
protocol, computing reconstructions of arbitrarily oriented slices in real time.

We note that the reconstruction step deviates slightly from the Noise2Inverse
reconstruction described in Equation~\eqref{eq:n2i-reconstruction}.
Rather than averaging separate reconstructions of each subset of the projection
data, Noise2Filter computes a reconstruction using the learned filters directly
from all data.
In the context of self-supervised learning, this technique has been observed to
yield improved results~\cite{batson-2019-noise2}.

\bigskip
\textbf{Noise2Filter summary}

The Noise2Filter method consists of three steps.
A summary of these steps, and specifically the computations performed,
is given below:
\begin{enumerate}
  \item [1.]
    \textbf{Data preparation} Compute the input and target training pairs from the measured
    projection data $\y$.
    Specifically, split the measured projection data in $\numsplits$ equal
    sub-datasets and compute the following for the ortho-slices:
    \begin{align}
        \fbp(\y_i, \h) & \text{ for }  i = 1, \ldots, \numsplits\\
        \fbp(\y_i, \ee_j) & \text{ for } i = 1, \ldots, \numsplits, j=1,\ldots, \numbases.
    \end{align}
    The computational effort of this step is equal to $3(\numbases + 1)$ FBP
    reconstructions of a 2D slice.

    \item [2.] \textbf{Training}
    Obtain a random sample of $\numtrain$ voxels on the ortho-slices for
    inclusion in the training set.
    Compute the optimal parameters $\theta^\star$ that minimizes the training
    objective with respect to the sampled voxels.
    Note that the training time depends on the size of the training set, which
    may be fixed independent of the object size.

    \item [3.] \textbf{Reconstruction}
    Using the computed parameters $\theta^\star$, compute an NN-FBP
    reconstruction for the desired 2D slices.
    Recall from Equation~\eqref{eq:mlp-is-nnfbp} that the computational cost of
    an NN-FBP reconstruction is equivalent to $\numfilters$ FBP reconstructions.
\end{enumerate}

\section{Experimental setup}\label{sec:experimental-setup}

In this section we discuss the setup of the experiments.
Specifically, we describe the data used in the experiments, the implementation
of NN-FBP and Noise2Filter, and the measures used to quantify these comparisons.

\subsection{Simulated data}\label{sec:data}
A phantom was generated by removing 100,000 randomly-placed non-overlapping
balls from a foam cylinder.
The \texttt{foam\_ct\_phantom} package~\cite{pelt-2018-improv-tomog} was used to
generate analytical projection images with $2 \times$ supersampling, were each
pixel's value is averaged over four equally-spaced rays through the pixel.
The result contains $1024$ equally-spaced projection images with
$512 \times 768$ pixels.

In each experiment, the simulated projection images were corrupted with Poisson
noise of various levels of intensity, by altering the incident photon count per
pixel $I_{0}$.
The average absorption of the sample was $10\%$.
Reconstructions without Poisson noise and with Poisson noise ($I_0=1000$) are shown in
Figure~\ref{fig:imagegrid}.

\subsection{NN-FBP and Noise2Filter}\label{sec:implementation}

Noise2Filter and NN-FBP benefit from a shared implementation.
Therefore, most almost all implementation details are the same.
As in the original NN-FBP implementation~\cite{pelt-2013-fast-tomog}, the number
of learned filters is set to $\numfilters=4$, the non-linear activation function
is the sigmoid, the exponential binning parameter is set to $2$, but the filters
are piece-wise linear --- rather than piece-wise constant --- as proposed
in~\cite{lagerwerf2020automated}.
Moreover, changes have been made to the shared implementation in order to
accelerate data preparation, training, and reconstruction.

In the data preparation step, reconstructions are computed of the ortho-slices
rather than the full volume.
These reconstructions are performed using the RECAST3D software
package~\cite{buurlage2018real}.

Some changes have been made to the training procedure.
As in the original implementation, the training objective is minimized using the
Levenberg-Marquadt algorithm (LMA), which requires that the data samples are
split into a training set and a validation set.
Compared to the original implementation, however, the number of training samples is
reduced from $10^6$ to $5\cdot 10^4$.
The effect of this reduction is discussed in \secref{sec:hyper_params}.
In addition, training is accelerated by performing computations on the graphics
processing unit (GPU) using PyTorch~\cite{paszke-2017-autom-pytor}.

Final reconstructions are computed using the RECAST3D software
package~\cite{buurlage2018real}.

\textbf{NN-FBP}
The free parameters for the NN-FBP method are trained and tested
on \emph{separate} tomographic datasets.
The training dataset consists of paired noisy and noiseless reconstructions.
Supervised training minimizes the training objective in
Equation~\eqref{eq:nnfbp-training}.

\textbf{Noise2Filter}
The Noise2Filter parameters are optimized using self-supervised
training on the noisy test dataset, rather than on a separate training dataset.
No noiseless reconstructions are necessary for training.
Depending on the training strategy (\emph{X:1} or \emph{1:X}), training minimizes either
Equation~\eqref{eq:n2f-x1} or \eqref{eq:n2f-1x}.

\subsection{Quantitative measures}

Reconstruction accuracy is quantified using the the \emph{Peak Signal-to-Noise
  Ratio} (PSNR) and the \emph{Structural Similarity} (SSIM)
index~\cite{wang-2004-image-qualit-asses} metrics.
Both metrics were computed with respect to the noiseless reconstructed images
and using a data range that was determined by the minimum and maximum intensity
of the noiseless reconstructed images.
If not otherwise mentioned, the reported metrics are the average of the metric
as computed on the three ortho-slices.

\section{Experiments \& Results}\label{sec:experiments-results}

We performed several experiments to evaluate the Noise2Filter method.
We provide a short summary below.

\textbf{Reconstruction accuracy} We compare Noise2Filter to supervised NN-FBP
training and several standard FBP improvement strategies in terms of
reconstruction accuracy.

\textbf{Hyperparameter analysis} Implementation choices in the design of the
Noise2Filter method are analyzed, including the number of training samples,
training strategy (\emph{X:1} or \emph{1:X}), and number of splits.

\textbf{Timing} An analysis of data preparation and reconstruction speed is
given.

\textbf{Experimental data} The method is applied to experimental data, including
a showcase that illustrates the potential for use in dynamic control.

\subsection{Reconstruction accuracy comparison}

In this section, we assess the reconstruction accuracy of the Noise2Filter
method.
We compare to other filter-based reconstruction techniques in terms of
reconstruction accuracy.
Specifically, we compare to a baseline FBP reconstruction (with a Ram-Lak filter) and FBP with standard
noise reduction techniques --- Gaussian filtering (FBP$_G$) and frequency
scaling (FBP$_{sc}$).
These two methods are discussed in more detail in \appref{sec:comp_meths}.
In addition, we compare to the NN-FBP, which is trained on a separate training
dataset with ground truth images.

The comparison is performed on the simulated foam dataset with varying levels of
Poisson noise.
The incident photon count $I_{0}$ was varied between $1000$ and $32,000$ in
powers of two.

For each of the methods, parameter selection was performed as follows.
For Noise2Filter, training was performed on the noisy test set.
For NN-FBP, training was performed on a separate training dataset.
For both methods, training was repeated 20 times to obtain statistics for the
PSNR and SSIM\@.
For Gaussian filtering and frequency scaling, the parameters maximizing the SSIM
on the test set were determined using a linear grid search.

The Noise2Filter method with the \emph{1:X} training strategy and $3$ splits is used.
We find that this yields consistent results at various noise levels.

\begin{figure}[!h]
  \centering
  \includegraphics[width=1\textwidth]{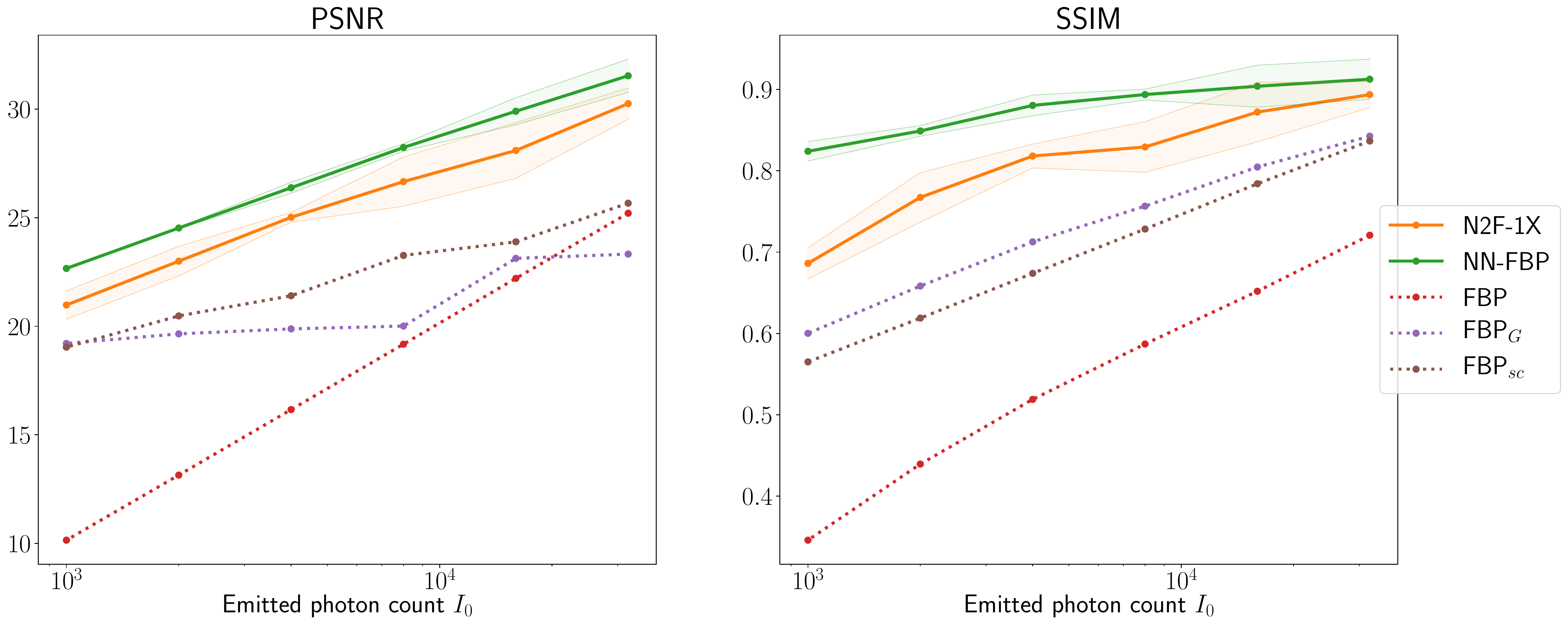}
  \caption{
    Reconstruction accuracy comparison of Noise2Filter (N2F-1X), NN-FBP, and FBP
    with Gaussian filtering, frequency scaling, and default filter.
    For varying noise levels, the average (line) and standard deviation (shaded region) over 20 trials of the PSNR and SSIM are reported.~\label{fig:QM_ortho}
  }
\end{figure}

\begin{figure}
  \centering
  \includegraphics[]{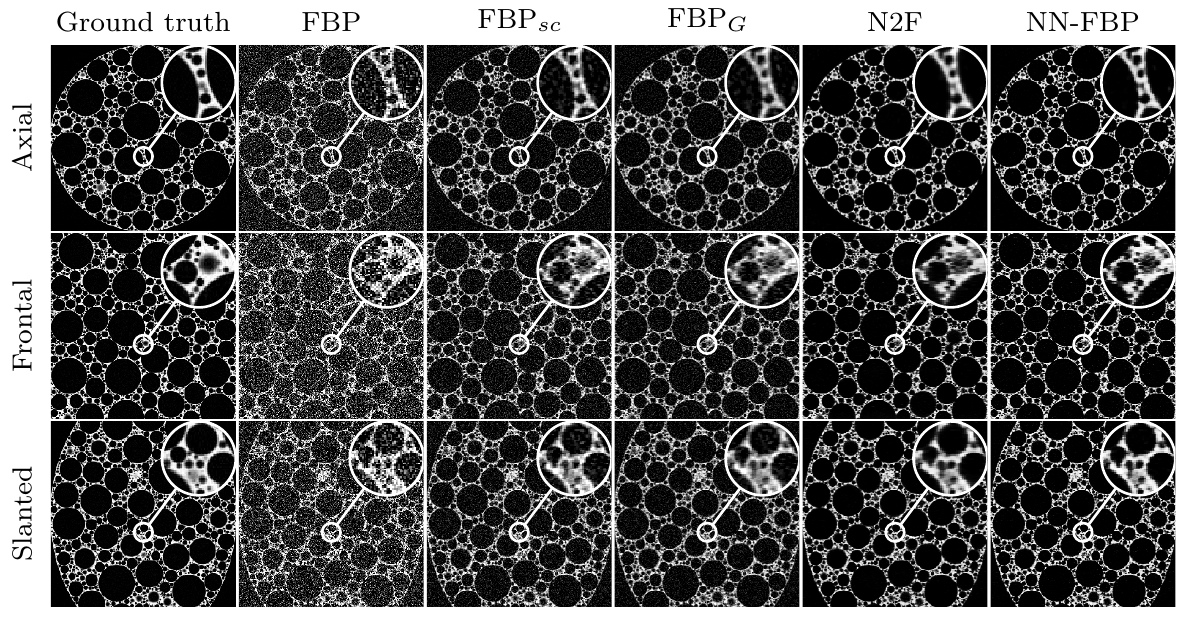}
  \caption{\label{fig:imagegrid} Results of noiseless reconstruction, FBP
    reconstruction, FBP with frequency scaling ($\fbp_{sc}$), FBP with Gaussian
    filtering ($\fbp_{G}$), Noise2Filter (N2F), and NN-FBP on a simulated foam
    phantom with photon count $I_{0} = 1000$.
    Results are shown on an axial, frontal, and $45^{\circ}$ slanted slice.
    The insets are zoomed by a factor of four.
  }
\end{figure}

The quantitative measures for the ortho-slices are shown in
\figref{fig:QM_ortho}.
For all noise levels, the Noise2Filter metrics are higher than FBP with 
frequency scaling or Gaussian filtering.
The NN-FBP method attains the best metrics, although the difference with
Noise2Filter decreases as the noise level decreases.
The difference in reconstruction accuracy is illustrated in
\figref{fig:imagegrid}, where the ground truth phantom and reconstructions for
all considered methods are shown for the incident photon count $I_0=1000$.
Notice that NN-FBP and Noise2Filter remove the noise in the voids, unlike the
FBP methods.

\subsection{Hyper parameter analysis}\label{sec:hyper_params}
We consider three hyper parameters for the N2F method: the number of samples
considered for training, the training strategy \emph{X:1} or \emph{1:X} and the
number of splits $\numsplits$ for the measured projection data.

\begin{figure}[!h]
  \centering
  \includegraphics[width=\textwidth]{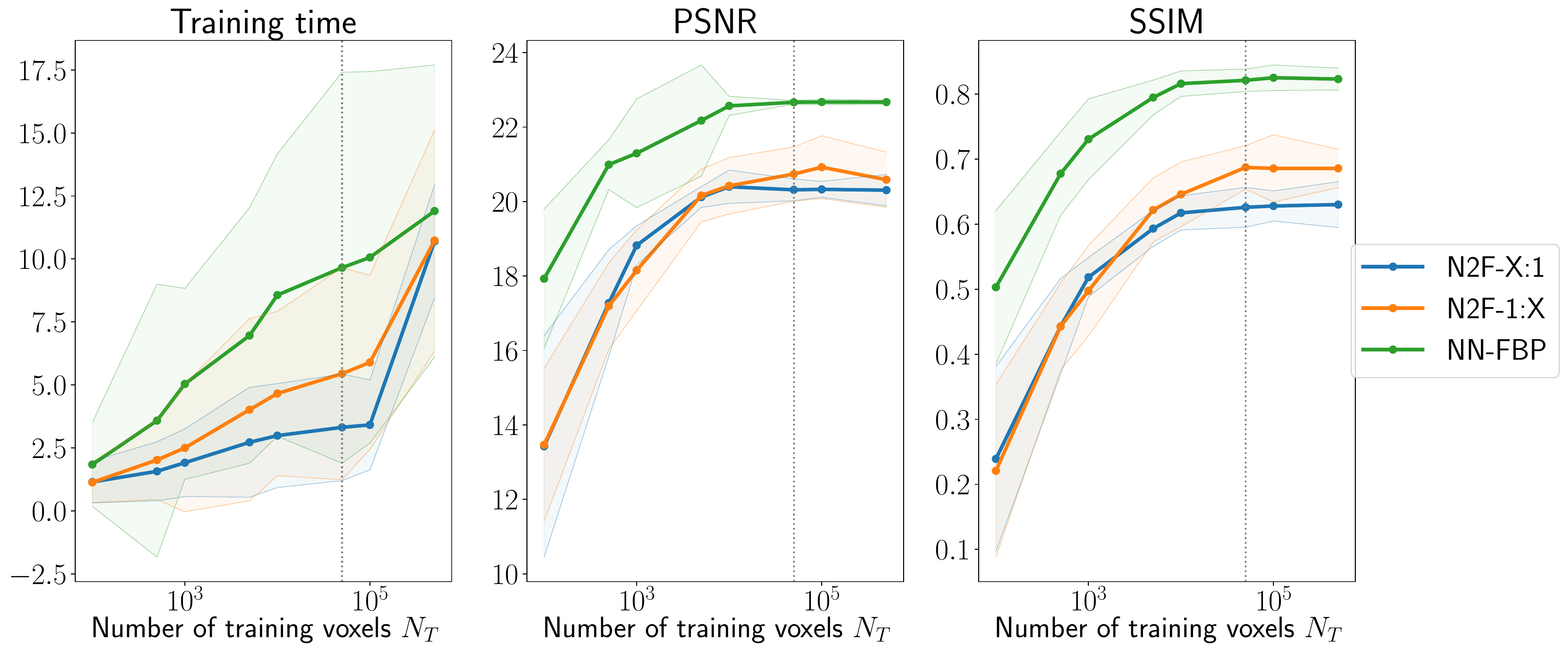}
  \caption{ Training time and reconstruction accuracy for varying amounts of
    training voxels $\numtrain$.
    The mean (line) and standard deviation (shaded region) over 50 trials are reported.
    For both NN-FBP and Noise2Filter, increasing $\numtrain$ yields diminishing
    returns in terms of PSNR and SSIM beyond $\numtrain=5 \cdot 10^{4}$, as
    indicated by the dashed line.~\label{fig:num_vox}
  }
\end{figure}

First, we analyzed the reconstruction accuracy as a function of $\numtrain$,
the number of training samples used in the training process.
Here, the number of validation samples is fixed to $10\%$ of the number of
training samples.
Noise was applied to the projection dataset equivalent to $I_{0}=1000$.
The results for this experiment are shown in \figref{fig:num_vox}.
We observe that increasing the number of voxels yields virtually no increase in
PSNR or SSIM beyond $\numtrain=5 \cdot 10^{4}$ voxels.

\begin{figure}[!h]
  \centering
  \includegraphics[width=\textwidth]{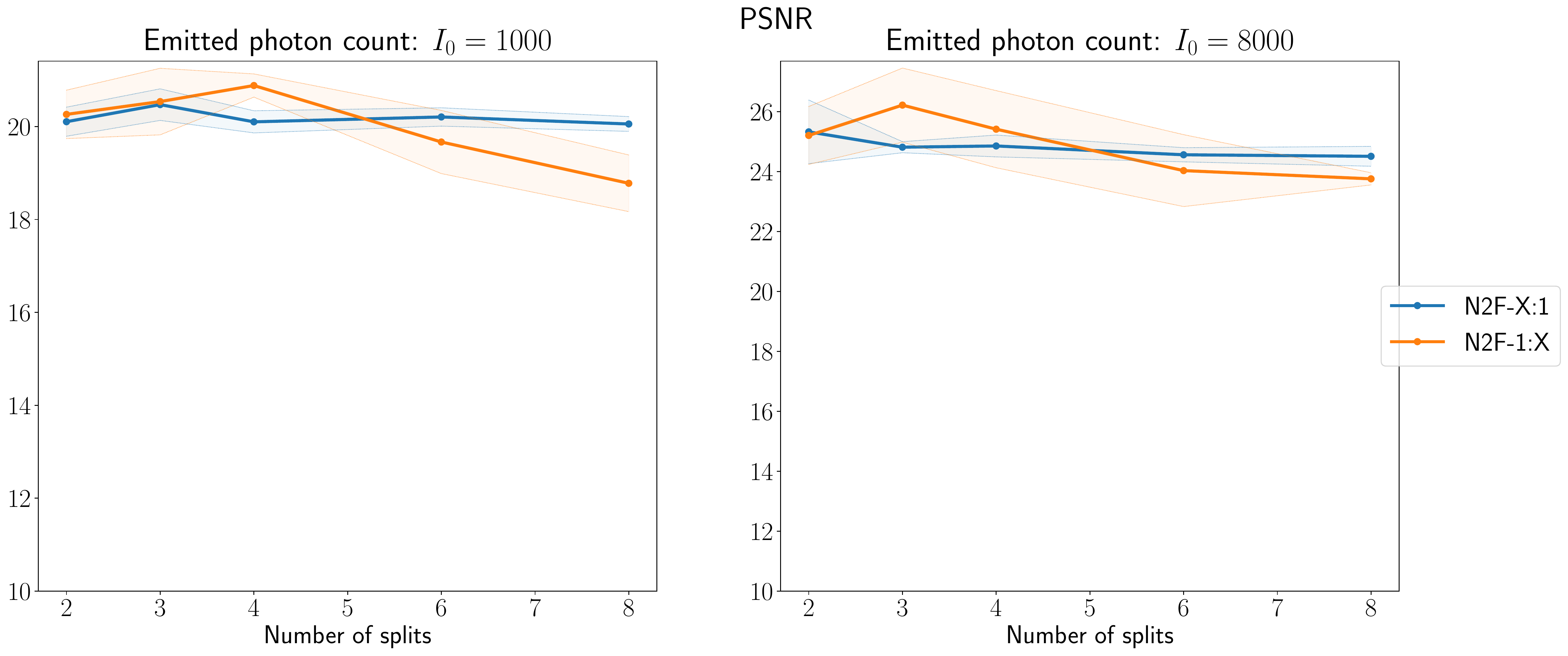}
  \includegraphics[width=\textwidth]{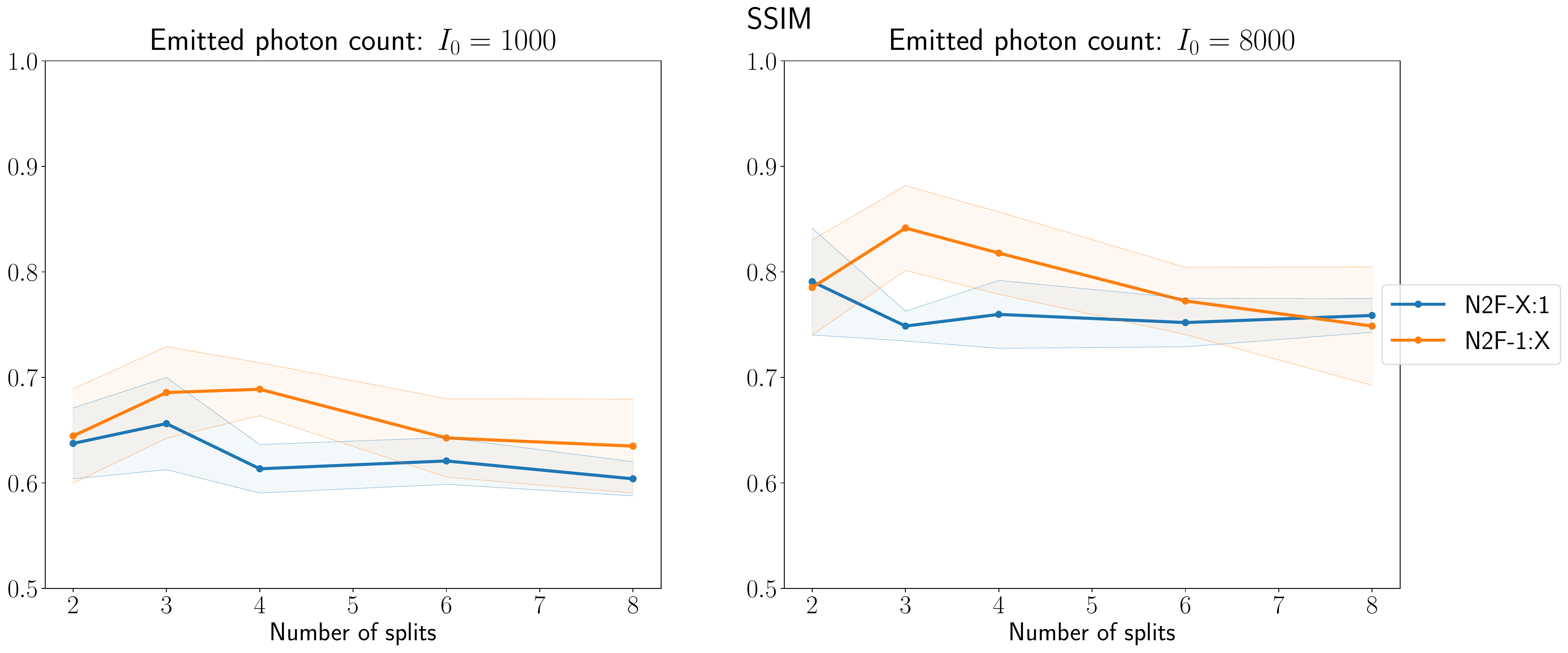}
  \caption{A comparison of Noise2Filter reconstruction accuracy for varying
    number of splits $\numsplits$ and training strategies X:1 and 1:X.
    Mean (line) and standard deviation (shaded region) over 20 trials of the PSNR and SSIM are plotted
    for noise levels $I_{0}=1000$, and $I_{0}= 8000$.~\label{fig:hyp_param}
  }
\end{figure}

Second, we compare the training strategies and the number of splits on the
simulated foam dataset for two noise levels, $I_{0}= 1000$ and $I_{0}=8000$.
For various values of the number of splits, 20 networks were trained and
used to reconstruct the projection data.
The average and standard deviation of the PSNR and SSIM are shown in
\figref{fig:hyp_param}.
For both noise levels we observe that the \emph{1{:}X} strategy with $3$ splits
obtains the best SSIM and close to the best PSNR.

\subsection{Timing comparison}

We give timings for the data preparation and training step of the Noise2Filter
method for several problem sizes.
The computations were performed on a server with 375 GB of RAM and made use of a
single Nvidia GeForce RTX 2080 Ti GPU (Nvidia, Santa Clara, CA, USA).

In \tabref{tab:time_tomosipo} we report the reconstruction times of one 2D slice
using the RECAST3D framework for standard FBP and the Noise2Filter method.
We see that Noise2Filter is roughly 4 times slower than standard FBP, which is
expected considering that we use $\numfilters=4$ learned filters.

\begin{table}
  \centering
      \begin{tabular}{rrrrrrr}
        \multicolumn{4}{c}{Data size} & \multicolumn{3}{|c}{Duration (seconds)}                                  \\
      \multicolumn{1}{c}{\# voxels}   & \multicolumn{1}{c}{\# pixels}    &
      \multicolumn{1}{c}{\# angles}   & \multicolumn{1}{c|}{$\numbases$} &
      \multicolumn{1}{c}{DP}      &
      \multicolumn{1}{c}{FBP}         & \multicolumn{1}{c}{N2F} \\
        \midrule
        $128^{3}$                     & $128 \times 192$                 & 256  & 10 & 0.34  & $0.003$ & $0.009$ \\
        $256^{3}$                     & $256 \times 384$                 & 512  & 11 & 1.34  & $0.006$ & $0.024$ \\
        $512^{3}$                     & $512 \times 768$                 & 1024 & 12 & 6.08  & $0.030$ & $0.114$ \\
        $1024^{3}$                    & $1024 \times 1536$               & 2048 & 13 & 44.00 & ---     & ---     \\
    \bottomrule
    \end{tabular}
    \caption{\label{tab:time_tomosipo} Benchmark results for the data
      preparation (DP) and reconstruction steps.
      FBP and Noise2Filter (N2F) reconstructions are performed on a single slice from
      filtered projection data.
      Due to memory constraints, some reconstructions not be performed, as
      indicated by a ---.
      }
\end{table}

\subsection{TomoBank dynamic dataset}

We consider two experiments with an experimental dynamic tomographic dataset,
consisting of $60$ scans at consecutive time steps.
First, we train Noise2Filter on the data from the first time step and use the
trained reconstruction method to compute reconstructions for later time steps.
This experiment aims to reveal the ability of Noise2Filter to generalize over
dynamics in time.
Second, we consider determining the correct center of rotation using
Noise2Filter.

The experimental data is taken from the public TomoBank
repository~\cite{de-2018-tomob} and was acquired at the TOMCAT beamline at the
Swiss Light Source (Paul Scherrer Institut, Switzerland).
In this experiment, sub-second X-ray tomographic microscopy was used to
investigate liquid water dynamics in a fuel cell during operation.
The experiment took less than 6 seconds, during which 60 scans were acquired.
A scan consists of 301 projections taken by a detector with $1100\times 1440$
detector pixels.

\begin{figure}[!h]
    \centering
    \includegraphics[width=1\textwidth]{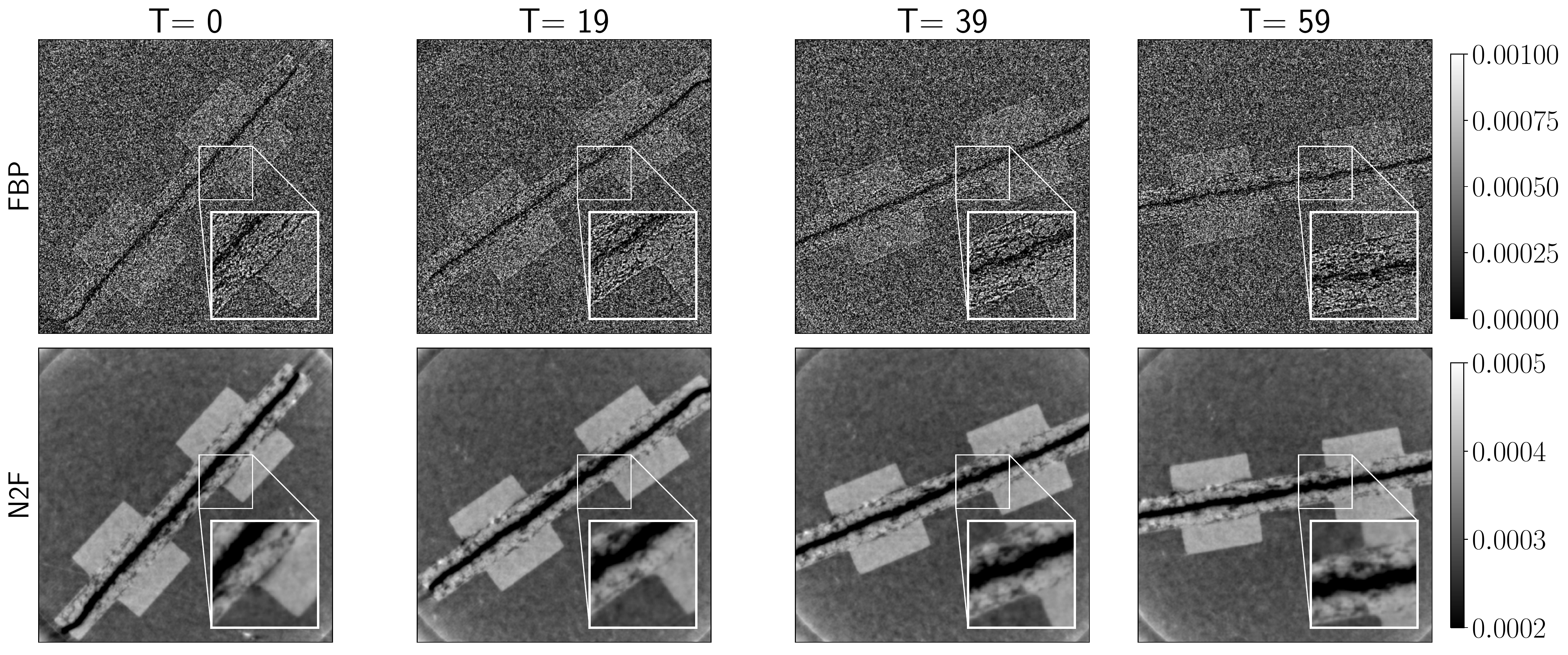}
    \caption{ Reconstruction of the fuel cell at various time steps using FBP
      and Noise2Filter (N2F).
      The Noise2Filter method was trained on the first time step and also used
      to reconstruct later time steps. The insets are zoomed by a factor two.\label{fig:fuel-cell}
    }
\end{figure}

First, we train a Noise2Filter network at the first time step $T=0$ and use this
network to evaluate all further time steps.
\figref{fig:fuel-cell} shows the results for this strategy for $T=0, 19, 39, 59$
and the FBP reconstructions at these time steps.
There is no visible deterioration of the reconstruction accuracy over time,
indicating that the trained network generalizes over the whole experiment.

\begin{figure}[!h]
    \centering
    \includegraphics[width=1\textwidth]{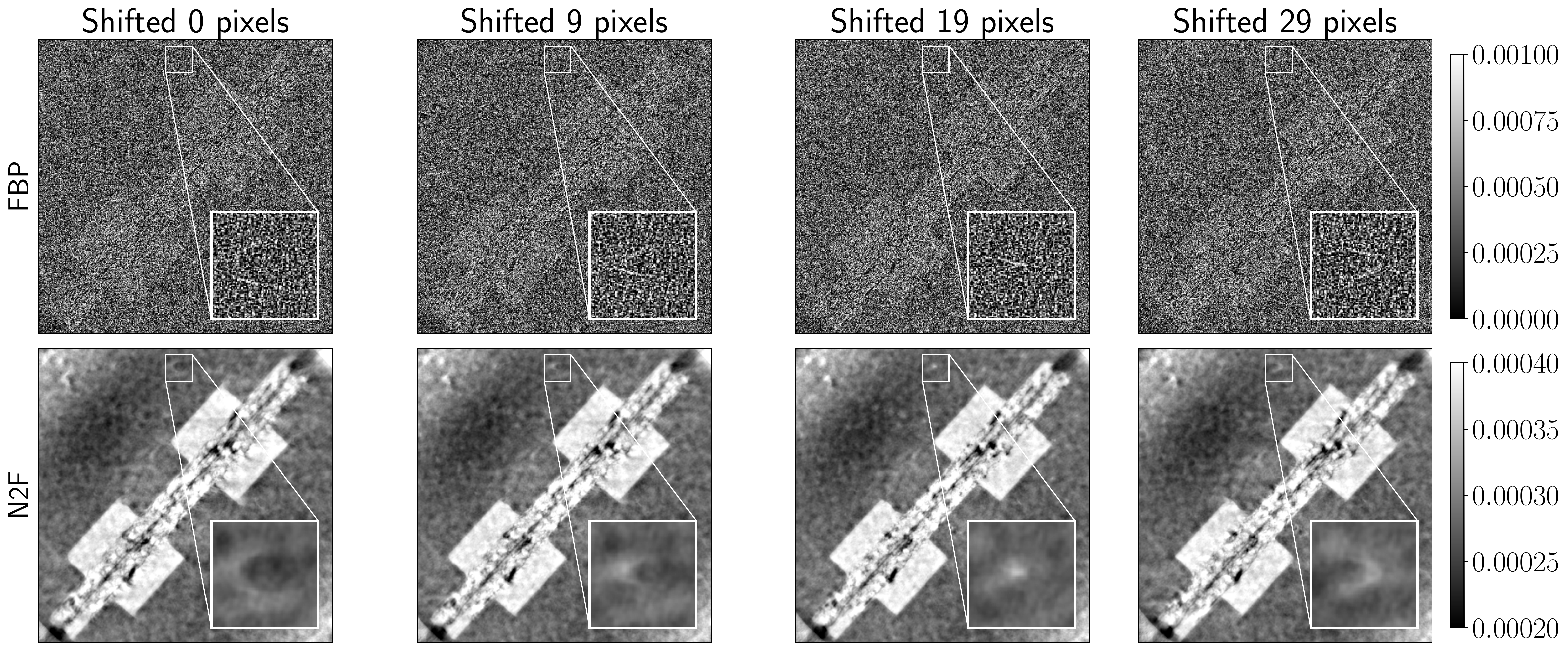}
    \caption{
      Reconstructions of a fuel cell at various centers of rotation using FBP
      and Noise2Filter (N2F).
      In the inset, a center of rotation artifact is highlighted, which
      disappears at a shift of $19$ pixels.
      The contrast in the Noise2Filter reconstruction is increased. The insets are zoomed by a factor four.\label{fig:cor-correction}
    }
\end{figure}

Second, we consider determining the correct center of rotation.
In the presence of noise, determining the correct center of rotation for a
dataset can be difficult and is often performed after acquiring the measured
projection data.
Using the tools developed in~\cite{vanrompay2020real}, the center of rotation
can be adapted interactively in real-time.
In \figref{fig:cor-correction} we show Noise2Filter and FBP reconstructions with
shifted centers of rotation at the first time step.
We note that no retraining was performed for Noise2Filter: the network
parameters were determined once using a shift of $0$ pixels.
In the FBP reconstructions, the center of rotation artifacts (half moons) are
difficult to discern.
In the Noise2Filter reconstruction, however, these artifacts are both clearly
visible, and visibly disappear at a shift of $19$ pixels, which coincides with
the reported center of rotation in~\cite{de-2018-tomob}.

\section{Conclusion}\label{sec:conclusion}
We have introduced Noise2Filter, a machine learning method for
denoising filter-based reconstruction that does not require any additional
training data beyond the acquired measurements.
We show that this self-supervised method improves reconstruction accuracy
compared to standard filter-based methods, and has limited loss of accuracy
compared to its supervised counterpart (NN-FBP).
The method exhibits sub-minute training times and reconstruction times in the order
of hundred milliseconds,
which demonstrates the potential for use in quasi-3D reconstruction for
real-time visualization of tomographic experiments.
In addition, we demonstrate that visual calibration of the center of rotation is
possible, which illustrates the potential of our method for use in the dynamic
control of tomographic experiments where noise is a challenge.

\section*{Acknowledgment}
The authors acknowledge financial support from the Dutch Research Council (NWO),
project number 639.073.506.
The authors have made use of the following additional software packages to
compute and visualize the experiments: Matplotlib and
scikit-image~\cite{hunter-2007-matpl,walt-2014-scikit-image}.


\bigskip
\appendix
\section{Standard FBP improvement strategies}\label{sec:comp_meths}

In addition to standard FBP and NN-FBP, the Noise2Filter method is compared to
two commonly used strategies to improve the reconstruction accuracy of the FBP
algorithm for noisy data~\cite{russo2017handbook}.

\subsection{Gaussian filtering}

In this strategy the standard filter $\h$ in the FBP algorithm is convolved with
a Gaussian filter $G_\sigma\in \R^{N_f}$ to smooth the noise in the
reconstructions, with $\sigma$ the standard deviation of the Gaussian.
The elements $j$ of the filter $G_\sigma$ are defined as follows:
\begin{align}
    (G_\sigma)_j = \tfrac{1}{\sigma \sqrt{2 \pi}}e^{-\tfrac{(j-N_f/2)^2}{2\sigma^2}},
\end{align}
resulting in the smoothed reconstruction
$\fbp_{G}(\y, \h, \sigma) = W^T(\y\ast (\h \ast G_\sigma))$.

\subsection{Frequency scaling}

This strategy removes the higher frequencies from the FBP reconstruction.
This is done by setting the frequencies above a threshold $f_{sc}$ in Fourier
domain of the filter $\h$ equal to zero and using this filter in the standard
FBP algorithm, obtaining $\fbp_{sc}(\y, \h_{sc}) = W^T(\y\ast \h_{sc})$.

For these strategies we optimized the choice of variable by computing reconstructions with a range of variables and taking the reconstruction with the highest SSIM.
\bibliographystyle{abbrv}
\bibliography{bibliography}

\end{document}